\documentclass[a4paper,12pt]{article}
\usepackage{amssymb}

\newtheorem{lemma}{Lemma}
\newtheorem{theorem}{Theorem}
\newtheorem{proposition}{Proposition}

\begin{document}

\author{
Dumitru B\u{a}leanu\\
\small{\c{C}ankaya University,}\\
\small{Department of Mathematics {\&} Computer Science,}\\
\small{\"{O}gretmenler Cad. 14 06530, Balgat -- Ankara, Turkey}\\
\small{e-mail address: dumitru@cankaya.edu.tr}\\
Octavian G. Mustafa\\
\small{University of Craiova, DAL,}\\
\small{Department of Mathematics {\&} Computer Science,}\\
\small{Tudor Vladimirescu 26, 200534 Craiova, Romania}\\
\small{e-mail address: octawian@yahoo.com}\\
Ravi P. Agarwal\\
\small{Florida Institute of Technology,}\\
\small{Department of Mathematical Sciences,}\\
\small{Melbourne, FL 32901, USA}\\
\small{e-mail address: agarwal@fit.edu}
}

\title{Asymptotic integration of $(1+\alpha)$-order fractional differential equations}
\date{}
\maketitle

\noindent{\bf Abstract} We establish the long-time asymptotic formula of solutions to the $(1+\alpha)$--order fractional differential equation ${}_{0}^{\>i}{\cal O}_{t}^{1+\alpha}x+a(t)x=0$, $t>0$, under some simple restrictions on the functional coefficient $a(t)$, where ${}_{0}^{\>i}{\cal O}_{t}^{1+\alpha}$ is one of the fractional differential operators ${}_{0}D_{t}^{\alpha}(x^{\prime})$, $({}_{0}D_{t}^{\alpha}x)^{\prime}={}_{0}D_{t}^{1+\alpha}x$ and ${}_{0}D_{t}^{\alpha}(tx^{\prime}-x)$. Here, ${}_{0}D_{t}^{\alpha}$ designates the Riemann-Liouville derivative of order $\alpha\in(0,1)$. The asymptotic formula reads as $[a+O(1)]\cdot x_{{\scriptstyle small}}+b\cdot x_{{\scriptstyle large}}$ as $t\rightarrow+\infty$ for given $a$, $b\in\mathbb{R}$, where $x_{{\scriptstyle small}}$ and $x_{{\scriptstyle large}}$ represent the eventually small and eventually large solutions that generate the solution space of the fractional differential equation ${}_{0}^{\>i}{\cal O}_{t}^{1+\alpha}x=0$, $t>0$.

\section{Introduction}
The present note continues our recent papers \cite{bma1,bma2,bma3} devoted to the fractional calculus variants of several fundamental results from the asymptotic integration theory of ordinary differential equations.

Let us consider the fractional differential equation (FDE) of order $1+\alpha$, with $\alpha\in(0,1)$, below
\begin{eqnarray}
{}_{0}^{\>i}{\cal O}_{t}^{1+\alpha}x+a(t)x=0,\quad t>0,\label{gen_fde0}
\end{eqnarray}
where the functional coefficient $a:[0,+\infty)\rightarrow\mathbb{R}$ is assumed continuous. The differential operator ${}_{0}^{\>i}{\cal O}_{t}^{1+\alpha}$ is a fractional version of the second order operator $\frac{d^{2}}{dt^{2}}$, built by taking into account the decompositions
\begin{eqnarray*}
x^{\prime\prime}=(x^{\prime})^{\prime},\quad tx^{\prime\prime}=(tx^{\prime}-x)^{\prime},\quad t>0,
\end{eqnarray*}
in the ring of smooth functions over $(0,+\infty)$.

To declare the operator ${}_{0}^{\>i}{\cal O}_{t}^{1+\alpha}$, denote by ${\cal RL}^{\alpha}((0,+\infty),\mathbb{R})$ the real linear space of all the functions $f\in C((0,+\infty),\mathbb{R})$ with $\lim\limits_{t\searrow0}[t^{1-\alpha}x(t)]\in\mathbb{R}$. Recall now the Rie\-mann-Liouville derivative of order $\alpha$ of the function $f\in {\cal RL}^{\alpha}((0,+\infty),\mathbb{R})$, namely
\begin{eqnarray*}
({}_{0}D_{t}^{\alpha}f)(t)=\frac{1}{\Gamma(1-\alpha)}\cdot\frac{d}{dt}\left[\int_{0}^{t}\frac{f(s)}{(t-s)^{\alpha}}ds\right],\quad t>0,
\end{eqnarray*} 
where $\Gamma$ stands for Euler's function Gamma, cf. \cite[p. 68]{podlubny}. If the function $f$ is at least absolutely continuous, see \cite[p. 35, Lemma 2.2]{SamkoKilbas}, then the derivative exists almost everywhere. Now, we introduce the quantities
\begin{eqnarray*}
{}_{0}^{\>1}{\cal O}_{t}^{1+\alpha}={}_{0}D_{t}^{\alpha}\circ\frac{d}{dt},\quad{}_{0}^{\>2}{\cal O}_{t}^{1+\alpha}=\frac{d}{dt}\circ{}_{0}D_{t}^{\alpha}
\end{eqnarray*}
and
\begin{eqnarray*}
{}_{0}^{\>3}{\cal O}_{t}^{1+\alpha}={}_{0}D_{t}^{\alpha}\circ\left(t\cdot\frac{d}{dt}-\mbox{id}_{{\cal RL}^{\alpha}((0,+\infty),\mathbb{R})}\right).
\end{eqnarray*}

The different factorisations \cite{ritt} of a fractional differential operator might lead to some interesting models in mathematical physics. We can mention that the fractional differential equations \cite{Trujillo,miller_ross,SamkoKilbas} are playing an important role in fluid dynamics, traffic model with fractional derivative, measurement of viscoelastic material properties, modeling of viscoplasticity, control theory, economy, nuclear magnetic resonance, mechanics, optics, signal processing and so on. Basically, the fractional differential equations are used to investigate the dynamics of the complex systems, the models based on these derivatives have given superior results as those based on the classical derivatives, see \cite[p. 305]{podlubny}, \cite{jiang,riewe, westerlund}.

Notice that the FDE
\begin{eqnarray}
{}_{0}^{\>i}{\cal O}_{t}^{1+\alpha}x=0,\quad t>0,\label{gen_fde1}
\end{eqnarray}
has a bidimensional solution space in ${\cal RL}^{\alpha}((0,+\infty),\mathbb{R})$ generated by the smooth functions
$1$ and $t^{\alpha}$ for $i=1$, $t^{\alpha-1}$ and $t^{\alpha}$ for $i=2$, and $t$ and $t^{\alpha-1}$ for $i=3$.

Regarding the equation (\ref{gen_fde0}) as {\it perturbation\/} of (\ref{gen_fde1}), one can ask {\it how close\/} a solution $x$ of (\ref{gen_fde0}) can get to the solution $a\cdot x_{{\scriptstyle small}}+b\cdot x_{{\scriptstyle large}}$ of (\ref{gen_fde1}), with $a$, $b\in\mathbb{R}$? Some simple restrictions on the functional coefficient $a(t)$ will be given next to ensure that an asymptotic formula for the general solution of each of the three FDEs exist similarly to the case of classical ordinary differential equations. In a loose manner, the formula reads as
\begin{eqnarray}
[a+O(1)]\cdot x_{{\scriptstyle small}}+b\cdot x_{{\scriptstyle large}}\quad \mbox{when }t\rightarrow+\infty.\label{as_form_gen}
\end{eqnarray}
Given the fact that the singular integral operators employed in our proofs resemble the integral operators from the two-point boundary value problems encountered in the theory of second order differential equations, see \cite{agarw}, we think that the Landau symbol $O(1)$ in our formula cannot be replaced with its counterpart $o(1)$ in the majority of circumstances.

\section{The case of ${}_{0}^{\>1}{\cal O}_{t}^{1+\alpha}$}

To establish (\ref{as_form_gen}), we introduce an integral operator acting within a complete metric space and prove that it is {\it a contraction\/} with respect to the space metric. The existence of its fixed point will follow then from the Contraction Principle and the solution based on the fixed point will obey the asymptotic formula.

We start with a formal derivation of the integral operator. Given $x\in C([0,+\infty),\mathbb{R})$ such that $x^{\prime}\in {\cal RL}^{\alpha}((0,+\infty),\mathbb{R})$, we integrate (\ref{gen_fde0}) over $[t,+\infty)$ to get
\begin{eqnarray*}
\frac{1}{\Gamma(1-\alpha)}\int_{0}^{t}\frac{x^{\prime}(s)}{(t-s)^{\alpha}}ds=x_{1}+\int_{t}^{+\infty}(ax)(s)ds,\quad t>0,
\end{eqnarray*}
where $x_{1}=\lim\limits_{t\rightarrow+\infty}\frac{1}{\Gamma(1-\alpha)}\int_{0}^{t}\frac{x^{\prime}(s)}{(t-s)^{\alpha}}ds\in\mathbb{R}$.

Further,
\begin{eqnarray*}
&&\frac{1}{\Gamma(1-\alpha)}\int_{0}^{t}\frac{1}{(t-s)^{1-\alpha}}\int_{0}^{s}\frac{x^{\prime}(u)}{(s-u)^{\alpha}}duds\\
&&=x_{1}\cdot\frac{t^{\alpha}}{\alpha}+\int_{0}^{t}\frac{1}{(t-s)^{1-\alpha}}\int_{s}^{+\infty}(ax)(u)duds.
\end{eqnarray*}

A Fubini-Tonelli argument, see \cite[p. 29]{SamkoKilbas}, leads to
\begin{eqnarray*}
&&\frac{1}{\Gamma(1-\alpha)}\int_{0}^{t}\frac{1}{(t-s)^{1-\alpha}}\int_{0}^{s}\frac{x^{\prime}(u)}{(s-u)^{\alpha}}duds\\
&&=\frac{1}{\Gamma(1-\alpha)}\int_{0}^{t}x^{\prime}(u)\int_{u}^{t}\frac{ds}{(t-s)^{1-\alpha}(s-u)^{\alpha}}du\\
&&=\frac{1}{\Gamma(1-\alpha)}\int_{0}^{t}x^{\prime}(u)\int_{0}^{1}\frac{dv}{(1-v)^{1-\alpha}v^{\alpha}}ds\\
&&=\frac{B(\alpha,1-\alpha)}{\Gamma(1-\alpha)}\int_{0}^{t}x^{\prime}(u)du,
\end{eqnarray*}
where $B$ is the Beta function, cf. \cite[p. 6]{podlubny}. Since $B(q,r)=\frac{\Gamma(q)\Gamma(r)}{\Gamma(q+r)}$ and $\Gamma(1+q)\allowbreak=q\Gamma(q)$, with $q$, $r\in(0,1)$, we obtain that
\begin{eqnarray*}
x(t)=x_{0}+\frac{x_1}{\Gamma(1+\alpha)}\cdot t^{\alpha}+\frac{1}{\Gamma(\alpha)}\int_{0}^{t}\frac{1}{(t-s)^{1-\alpha}}\int_{s}^{+\infty}(ax)(\tau)d\tau ds,
\end{eqnarray*}
with $x(0)=x_{0}\in\mathbb{R}$.

Taking $a=x_{0}$, $b=\frac{x_1}{\Gamma(1+\alpha)}$, with $a^{2}+b^{2}>0$, the integral operator reads as
\begin{eqnarray}
{\cal T}(x)(t)=a+bt^{\alpha}+\frac{1}{\Gamma(\alpha)}\int_{0}^{t}\frac{1}{(t-s)^{1-\alpha}}\int_{s}^{+\infty}(ax)(\tau)d\tau ds,\quad t>0.\label{op1}
\end{eqnarray}

\begin{theorem}\label{teor1}
Assume that there exists $T>0$ such that $\int_{T}^{+\infty}s^{1+\alpha}\vert a(s)\vert ds<+\infty$ and
\begin{eqnarray*}
\frac{\max\{1,T^{\alpha}\}}{\Gamma(1+\alpha)}\left[\int_{0}^{T}\vert a(s)\vert ds+\int_{T}^{+\infty}s^{\alpha}\vert a(s)\vert ds\right]=k<1.
\end{eqnarray*}
Then the FDE (\ref{gen_fde0}) for $i=1$ has a solution $x\in C([0,+\infty),\mathbb{R})$ with the asymptotic formula
\begin{eqnarray}
x(t)=a+bt^{\alpha}+O(t^{\alpha-1})=a+bt^{\alpha}+o(1)\quad\mbox{when }t\rightarrow+\infty.\label{as_form_op1}
\end{eqnarray}
In particular, $O(1)$ can be replaced with $o(1)$ in (\ref{as_form_gen}).
\end{theorem}

{\bf Proof.} Let $X$ be the set of all the functions $x\in C([0,+\infty),\mathbb{R})$ with $\sup\limits_{t\geq T}\frac{\vert x(t)\vert}{t^{\alpha}}<+\infty$ and $d$ the following metric
\begin{eqnarray*}
d(x_{1},x_{2})=\max\left\{\Vert x_{1}-x_{2}\Vert_{L^{\infty}(0,T)},\sup\limits_{t\geq T}\frac{\vert x_{1}(t)-x_{2}(t)\vert}{t^{\alpha}}\right\},\quad x_{1},x_{2}\in X.
\end{eqnarray*} 
Obviously, ${\cal M}=(X,d)$ is a complete metric space.

Notice that
\begin{eqnarray*}
\int_{0}^{+\infty}s^{j}\vert ax\vert(s)ds&\leq&\left[\int_{0}^{T}s^{j}\vert a(s)\vert ds+\int_{T}^{+\infty}s^{j+\alpha}\vert a(s)\vert ds\right]d(x,0)\\
&=&C(j)\cdot d(x,0),
\end{eqnarray*}
where $j\in\{0,1\}$, for every $x\in {\cal M}$.

Introduce the operator ${\cal T}:{\cal M}\rightarrow C([0,+\infty),\mathbb{R})$ with the formula (\ref{op1}). We have the estimates
\begin{eqnarray*}
\vert {\cal T}(x)(t)\vert&\leq& \vert a\vert+\vert b\vert t^{\alpha}+\frac{1}{\Gamma(\alpha)}\int_{0}^{t}\frac{ds}{(t-s)^{1-\alpha}}\cdot\int_{0}^{+\infty}\vert ax\vert(s)ds\\
&\leq&\vert a\vert+T^{\alpha}\left[\vert b\vert+\frac{C(0)}{\Gamma(1+\alpha)}\cdot d(x,0)\right],\quad t\in[0,T],
\end{eqnarray*}
and
\begin{eqnarray*}
\vert {\cal T}(x)(t)\vert&\leq& t^{\alpha}\left[\frac{\vert a\vert+T^{\alpha}\vert b\vert}{T^{\alpha}}+\frac{C(0)}{\Gamma(1+\alpha)}\cdot d(x,0)\right],\quad t\geq T,
\end{eqnarray*}
which imply that ${\cal T}(x)\in{\cal M}$ and
\begin{eqnarray*}
&&d({\cal T}(x),0)\\
&&\leq\max\left\{1,\frac{1}{T^{\alpha}}\right\}(\vert a\vert+T^{\alpha}\vert b\vert)+\max\{1,T^{\alpha}\}\frac{C(0)}{\Gamma(1+\alpha)}\cdot d(x,0),
\end{eqnarray*}
where $x\in{\cal M}$.

We also have
\begin{eqnarray*}
d({\cal T}(x_{1}),{\cal T}(x_{2}))\leq\frac{\max\{1,T^{\alpha}\}}{\Gamma(1+\alpha)}C(0)\cdot d(x_{1},x_{2}),\quad x_{1},x_{2}\in{\cal M},
\end{eqnarray*}
which means that ${\cal T}:{\cal M}\rightarrow{\cal M}$ is a contraction of coefficient $k$.

Let $x_{0}\in{\cal M}$ be its fixed point. Following verbatim the computations from \cite[eqs. (10), (16)]{bma2}, we have the estimates
\begin{eqnarray*}
&&\int_{0}^{t}\frac{1}{(t-s)^{1-\alpha}}\int_{s}^{+\infty}\vert ax_{0}\vert(\tau)d\tau ds\\
&&=\int_{0}^{t}\vert ax_{0}\vert(\tau)\frac{t^{\alpha}-(t-\tau)^{\alpha}}{\alpha}d\tau+\frac{t^{\alpha}}{\alpha}\int_{t}^{+\infty}\vert ax_{0}\vert ds\\
&&\leq\frac{t^{\alpha}}{\alpha}\left[\int_{0}^{t}\vert ax_{0}\vert(\tau)\cdot\frac{\tau}{t}d\tau+\frac{1}{t}\int_{t}^{+\infty}s\vert ax_{0}\vert(s)ds\right]\\
&&\leq\frac{2C(1)}{\alpha}d(x_{0},0)\cdot t^{\alpha-1}=O(t^{\alpha-1})\quad\mbox{when }t\rightarrow+\infty.
\end{eqnarray*}

Finally,
\begin{eqnarray*}
x_{0}(t)={\cal T}(x_{0})(t)=a+bt^{\alpha}+O(t^{\alpha-1})\quad\mbox{when }t\rightarrow+\infty.
\end{eqnarray*}

The proof is complete. $\square$

A particular case of (\ref{as_form_op1}) has been undertaken in \cite{bma1}, namely the case when $a=1$, $b=0$. We asked there if, similarly to the circumstances of ordinary differential equations \cite[Section 7]{agarw}, the solution $x_0$ from Theorem \ref{teor1} would have the (powerful) asymptotic behavior
\begin{eqnarray*}
x_{0}(t)=1+o(1)\quad\mbox{as }t\rightarrow+\infty,\qquad x^{\prime}\in(L^{1}\cap L^{\infty})((0,+\infty),\mathbb{R}).
\end{eqnarray*}
We also noticed that, most probably, to get such a result one must look for a {\it sign-changing\/} functional coefficient $a(t)$, see \cite[Section 3]{bma1}. 

In the remaining of the present section we shall discuss the issue of {\lq\lq}$x^{\prime}\in L^{1}${\rq\rq} and conclude that this can happen (eventually) in very restricted conditions.

\begin{lemma}\label{lem1}
Assume that $a\in (C\cap L^{\infty})([0,+\infty),\mathbb{R})$ verifies the hypotheses from \cite{bma1}: it has a unique zero $t_{0}>0$, $\int_{0}^{+\infty}a(s)ds=0$, $\int_{0}^{+\infty}s\vert a(s)\vert ds<+\infty$ and $B\in (L^{1}\cap L^{\infty})([0,+\infty),\mathbb{R})$, where $B(t)=t^{\alpha}\Vert a\Vert_{L^{\infty}(t,+\infty)}$ for all $t\geq0$. Then, introducing the quantity $C(t)=\int_{0}^{t}\frac{a(s)}{(t-s)^{1-\alpha}}ds$, $t\geq0$, we have
\begin{eqnarray}
\int_{0}^{+\infty}\vert C(t)\vert dt+\sup\limits_{t\geq0}\vert C(t)\vert<+\infty.\label{intermed1}
\end{eqnarray}
If $B^{*}\in L^{1}([0,+\infty),\mathbb{R})$, where $B^{*}(t)=\sup\limits_{s\geq t}B(s)$ for all $t\geq0$, then
\begin{eqnarray}
\int_{0}^{+\infty}C^{*}(t)dt<+\infty,\quad C^{*}(t)=\sup\limits_{s\geq t}\vert C(s)\vert,\thinspace t\geq0.\label{intermed0}
\end{eqnarray}
If $\int_{0}^{+\infty}s^{1+\alpha}\vert a(s)\vert ds+\int_{0}^{+\infty}\Vert B\Vert_{L^{2}(t,+\infty)}dt<+\infty$ then we also have
\begin{eqnarray}
\int_{0}^{+\infty}\Vert C\Vert_{L^{2}(u,+\infty)}du=\int_{0}^{+\infty}\left(\int_{u}^{+\infty}\vert C(t)\vert^{2}dt\right)^{\frac{1}{2}}du<+\infty.\label{intermed2}
\end{eqnarray}
\end{lemma}

{\bf Proof.} As in \cite{bma1}, for $t>0$, the following estimates are valid
\begin{eqnarray}
\vert C(2t)\vert\leq\frac{B(t)}{\alpha}+\left\vert\int_{0}^{t}\frac{a(s)}{(2t-s)^{1-\alpha}}ds\right\vert\label{intermed3}
\end{eqnarray}
and
\begin{eqnarray}
&&\int_{0}^{t}\frac{a(s)}{(2t-s)^{1-\alpha}}ds\nonumber\\
&&=t^{\alpha-1}\int_{0}^{t}a(s)ds-(1-\alpha)\int_{0}^{t}\frac{1}{(2t-s)^{2-\alpha}}\int_{0}^{s}a(\tau)d\tau ds\label{intermed4}\\
&&=-t^{\alpha-1}\int_{t}^{+\infty}a(s)ds+(1-\alpha)\int_{0}^{t}\frac{1}{(2t-s)^{2-\alpha}}\int_{s}^{+\infty}a(\tau)d\tau ds.\label{intermed5}
\end{eqnarray}

Since $B\in L^{1}\cap L^{\infty}$, it is obvious that $B\in L^{2}$, so we shall focus on the second member from the right part of (\ref{intermed3}). By means of (\ref{intermed4}), we get
\begin{eqnarray*}
&&\left\vert t^{\alpha-1}\int_{0}^{t}a(s)ds\right\vert+\left\vert\int_{0}^{t}\frac{1}{(2t-s)^{2-\alpha}}\int_{0}^{s}a(\tau)d\tau ds\right\vert\\
&&\leq t^{\alpha}\Vert a\Vert_{L^{\infty}}+\frac{1}{t^{2-\alpha}}\int_{0}^{t}(s\cdot\Vert a\Vert_{L^{\infty}})ds=\frac{3}{2}\Vert a\Vert_{L^{\infty}}\cdot t^{\alpha},
\end{eqnarray*}
which leads to $C\in (L^{1}\cap L^{\infty})([0,T_{0}],\mathbb{R})$, where $T_{0}=\max\{1,t_{0}\}$. Further, via (\ref{intermed5}), 
\begin{eqnarray*}
D(t)&=&\left\vert\int_{0}^{t}\frac{a(s)}{(2t-s)^{1-\alpha}}ds\right\vert\\
&\leq& t^{\alpha-1}\int_{t}^{+\infty}\vert a(s)\vert ds+t^{\alpha-2}\int_{t}^{+\infty}s\vert a(s)\vert ds\\
&\leq&\frac{2}{t^{2-\alpha}}\int_{t}^{+\infty}s\vert a(s)\vert ds,\quad t\geq T_{0},
\end{eqnarray*}
and so $C\in (L^{1}\cap L^{\infty})([T_{0},+\infty),\mathbb{R})$. The estimate (\ref{intermed1}) has been obtained. As a byproduct, $C\in L^{2}([0,+\infty),\mathbb{R})$.

To prove (\ref{intermed0}), introduce $D^{*}(t)=\sup\limits_{s\geq t}D(s)$ for all $t\geq0$. We rely on the estimates
\begin{eqnarray*}
D^{*}(t)\leq\frac{3}{2}\Vert a\Vert_{L^{\infty}}\cdot t^{\alpha},\quad t\in[0,T_{0}],
\end{eqnarray*}
and
\begin{eqnarray*}
\int_{T_0}^{+\infty}D^{*}(t)dt&\leq&2\int_{T_0}^{+\infty}\frac{ds}{s^{2-\alpha}}\cdot\int_{T_0}^{+\infty}\tau\vert a(\tau)\vert d\tau\\
&=&\frac{2T_{0}^{\alpha-1}}{(1-\alpha)}\int_{T_0}^{+\infty}\tau\vert a(\tau)\vert d\tau,
\end{eqnarray*}
since the mapping $t\mapsto t^{\alpha-2}\int_{t}^{+\infty}s\vert a(s)\vert ds$ is monotone non-increasing in $[T_{0},+\infty)$.

For the third part, notice that
\begin{eqnarray*}
D(t)\leq\frac{2}{t^{2}}\int_{t}^{+\infty}s^{1+\alpha}\vert a(s)\vert ds,\quad t\geq T_{0},
\end{eqnarray*}
and
\begin{eqnarray*}
&&\left(\int_{2u}^{+\infty}\vert C(2t)\vert^{2}dt\right)^{\frac{1}{2}}\\
&&\leq\frac{1}{\alpha}\cdot\Vert B\Vert_{L^{2}(2u,+\infty)}+\left(\int_{2u}^{+\infty}\frac{dt}{t^{4}}\right)^{\frac{1}{2}}\cdot2\int_{2u}^{+\infty}s^{1+\alpha}\vert a(s)\vert ds\\
&&\leq\alpha^{-1}\Vert B\Vert_{L^{2}(2u,+\infty)}+\frac{u^{-\frac{3}{2}}}{\sqrt{6}}\int_{0}^{+\infty}s^{1+\alpha}\vert a(s)\vert ds,\quad u\geq T_{0}.
\end{eqnarray*}
We have obtained that $\int_{T_0}^{+\infty}\left(\int_{2u}^{+\infty}\vert C(2t)\vert^{2}dt\right)^{\frac{1}{2}}du<+\infty$.

Finally,
\begin{eqnarray*}
\int_{0}^{T_0}\left(\int_{2u}^{+\infty}\vert C(2t)\vert^{2}dt\right)^{\frac{1}{2}}du&=&\frac{1}{\sqrt{2}}\int_{0}^{T_0}\left(\int_{4u}^{+\infty}\vert C(v)\vert^{2}dv\right)^{\frac{1}{2}}du\\
&\leq&\frac{T_0}{\sqrt{2}}\Vert C\Vert_{L^2(0,+\infty)}.
\end{eqnarray*}

The proof is complete. $\square$

\begin{lemma}\label{lem2}
Assume that the function $C$ from Lemma \ref{lem1} satisfies the restrictions (\ref{intermed1}), (\ref{intermed0}) and (\ref{intermed2}) and either
\begin{eqnarray*}
\Vert C\Vert_{L^{\infty}}+2\Vert C^{*}\Vert_{L^1}=k_{1}<1
\end{eqnarray*}
or
\begin{eqnarray*}
2\Vert C^{*}\Vert_{L^{1}}<1,\quad\max\left\{\Vert C\Vert_{L^{\infty}}+\Vert C\Vert_{L^2},\Vert C\Vert_{L^{1}}+\Vert E\Vert_{L^1}\right\}=k_{2}<1,
\end{eqnarray*}
where $E(t)=\Vert C\Vert_{L^{2}(t,+\infty)}$ for all $t\geq0$. Then there exists a function $y\in (C\cap L^{1}\cap L^{\infty})([0,+\infty),\allowbreak\mathbb{R})$ such that
\begin{eqnarray}
y(t)=-C(t)\left(1-\int_{t}^{+\infty}y(s)ds\right)-\int_{t}^{+\infty}(Cy)(s)ds,\quad t\geq0.\label{int_y_0}
\end{eqnarray}
\end{lemma}

{\bf Proof.} Set the number $\gamma>1$ such that
\begin{eqnarray*}
1+2\gamma\int_{0}^{+\infty}C^{*}(s)ds<\gamma.
\end{eqnarray*}

Introduce the set $Y$ of all the functions $y\in C([0,+\infty),\mathbb{R})$ such that $\vert y(t)\vert\leq\gamma\cdot C^{*}(t)$, $t\geq0$, and the metric $d$ with the formula
\begin{eqnarray*}
d(y_{1},y_{2})=\max\{\Vert y_{1}-y_{2}\Vert_{L^{\infty}(0,+\infty)},\Vert y_{1}-y_{2}\Vert_{L^{1}(0,+\infty)}\},\quad y_{1},y_{2}\in Y.
\end{eqnarray*}
Using the Dominated Convergence Theorem, we deduce that the metric space ${\cal N}=(Y,d)$ is complete.

Consider the integral operator ${\cal T}:{\cal N}\rightarrow C([0,+\infty),\mathbb{R})$ given by the right-hand member of (\ref{int_y_0}). The following estimates
\begin{eqnarray*}
\vert {\cal T}(y)(t)\vert&\leq&\vert C(t)\vert(1+\Vert y\Vert_{L^1})+C^{*}(t)\int_{t}^{+\infty}\vert y(s)\vert ds\\
&\leq&C^{*}(t)(1+2\Vert y\Vert_{L^1})\leq C^{*}(t)\cdot\left(1+2\gamma\int_{0}^{+\infty}C^{*}(s)ds\right)\\
&\leq&\gamma\cdot C^{*}(t),\quad t\geq0,
\end{eqnarray*}
show that ${\cal T}:{\cal N}\rightarrow{\cal N}$ is well-defined.

Now, we have
\begin{eqnarray*}
\vert{\cal T}(y_{1})(t)-{\cal T}(y_{2})(t)\vert&\leq& C^{*}(t)\Vert y_{1}-y_{2}\Vert_{L^1}+\int_{0}^{+\infty}\vert C(s)\vert ds\cdot\Vert y_{1}-y_{2}\Vert_{L^{\infty}}\\
&\leq&(\Vert C\Vert_{L^{\infty}}+\Vert C\Vert_{L^1})\cdot d(y_{1},y_{2}),
\end{eqnarray*}
by noticing that $C^{*}(0)=\Vert C\Vert_{L^{\infty}(0,+\infty)}$, and also
\begin{eqnarray*}
\int_{t}^{+\infty}\vert{\cal T}(y_{1})(s)-{\cal T}(y_{2})(s)\vert ds&\leq&\int_{t}^{+\infty}(\vert C(s)\vert\cdot \Vert y_{1}-y_{2}\Vert_{L^{1}}) ds\\
&+&\int_{t}^{+\infty}C^{*}(s)\int_{s}^{+\infty}\vert y_{1}(\tau)-y_{2}(\tau)\vert d\tau ds\\
&\leq&2\int_{0}^{+\infty}C^{*}(s)ds\cdot d(y_{1},y_{2}),\quad t\geq0,
\end{eqnarray*}
which lead to
\begin{eqnarray*}
d({\cal T}(y_{1}),{\cal T}(y_{2}))&\leq&\max\left\{\Vert C\Vert_{L^{\infty}}+\Vert C\Vert_{L^1},2\Vert C^{*}\Vert_{L^1}\right\}\cdot d(y_{1},y_{2})\\
&\leq& k_{1}d(y_{1},y_{2}),
\end{eqnarray*}
where $y_{1}$, $y_{2}\in{\cal N}$. 

Notice that we haven't employed (\ref{intermed2}). To do so, let us use different estimates, namely
\begin{eqnarray*}
\vert{\cal T}(y_{1})(t)-{\cal T}(y_{2})(t)\vert&\leq&\vert C(t)\vert\cdot\Vert y_{1}-y_{2}\Vert_{L^1}\\
&+&\left[\int_{t}^{+\infty}\vert C(s)\vert^{2}ds\right]^{\frac{1}{2}}\cdot\left[\int_{t}^{+\infty}\vert y_{1}(s)-y_{2}(s)\vert^{2}ds\right]^{\frac{1}{2}}
\end{eqnarray*}
and
\begin{eqnarray*}
\int_{t}^{+\infty}\vert y_{1}(s)-y_{2}(s)\vert^{2}ds&\leq&\sup\limits_{\tau\geq0}\vert y_{1}(\tau)-y_{2}(\tau)\vert\cdot\int_{t}^{+\infty}\vert y_{1}(s)-y_{2}(s)\vert ds\\
&\leq&[d(y_{1},y_{2})]^{2},\quad t\geq0.
\end{eqnarray*}
They imply
\begin{eqnarray*}
\vert{\cal T}(y_{1})(t)-{\cal T}(y_{2})(t)\vert&\leq&\left[\vert C(t)\vert+\left(\int_{t}^{+\infty}\vert C(s)\vert^{2}ds\right)^{\frac{1}{2}}\right]\cdot d(y_{1},y_{2})\\
&\leq&(\Vert C\Vert_{L^{\infty}}+\Vert C\Vert_{L^2})d(y_{1},y_{2})
\end{eqnarray*}
and
\begin{eqnarray*}
\int_{t}^{+\infty}\vert{\cal T}(y_{1})(s)-{\cal T}(y_{2})(s)\vert ds&\leq&\left[\Vert C\Vert_{L^{1}}+\int_{0}^{+\infty}\left(\int_{t}^{+\infty}\vert C(s)\vert^{2}ds\right)^{\frac{1}{2}}dt\right]\\
&\times&d(y_{1},y_{2}),
\end{eqnarray*}
thus leading to
\begin{eqnarray*}
d({\cal T}(y_{1}),{\cal T}(y_{2}))&\leq&\max\left\{\Vert C\Vert_{L^{\infty}}+\Vert C\Vert_{L^2},\Vert C\Vert_{L^{1}}+\Vert E\Vert_{L^1}\right\}\cdot d(y_{1},y_{2})\\
&\leq&k_{2}d(y_{1},y_{2}),
\end{eqnarray*}
where $y_{1}$, $y_{2}\in{\cal N}$. 

The operator ${\cal T}:{\cal N}\rightarrow{\cal N}$ being a contraction, its fixed point $y_{0}$ is the solution of (\ref{int_y_0}) we are looking for. The proof is complete. $\square$

\begin{proposition}\label{prop1}
Let $y\in C([0,+\infty),\mathbb{R})$ be the solution of (\ref{int_y_0}) from Lemma \ref{lem2}. If $y(0)=0$ then the function $x\in C^{1}([0,+\infty),\mathbb{R})$ with the formula $x(t)=1-\int_{t}^{+\infty}y(s)ds$ for all $t\geq0$ is a solution of the FDE (\ref{gen_fde0}) for $i=1$ which satisfies the restrictions
\begin{eqnarray*}
x(t)=1+o(1)\quad\mbox{as }t\rightarrow+\infty,\qquad x^{\prime}\in(L^{1}\cap L^{\infty})([0,+\infty),\mathbb{R}).
\end{eqnarray*}
\end{proposition}

{\bf Proof.} Following \cite{bma1}, the function $x$ verifies the identity
\begin{eqnarray}
y(t)&=&-\frac{1}{\Gamma(\alpha)}\int_{0}^{t}\frac{a(s)x(s)}{(t-s)^{1-\alpha}}ds=-\frac{1}{\Gamma(\alpha)}\int_{0}^{t}\frac{a(s)}{(t-s)^{1-\alpha}}ds\nonumber\\
&+&\frac{1}{\Gamma(\alpha)}\int_{0}^{t}\frac{a(s)}{(t-s)^{1-\alpha}}\left(\int_{s}^{t}+\int_{t}^{+\infty}\right)y(\tau)d\tau ds\nonumber\\
&=&-C(t)+\int_{0}^{t}y(\tau)C(\tau)d\tau+C(t)\int_{t}^{+\infty}y(\tau)d\tau\nonumber\\
&=&-C(t)\left(1-\int_{t}^{+\infty}y(s)ds\right)+\int_{0}^{t}(Cy)(s)ds,\quad t\geq0.\label{l1_issue_0}
\end{eqnarray}
We have rescaled $C$ as $C(t)=\frac{1}{\Gamma(\alpha)}\int_{0}^{t}\frac{a(s)}{(t-s)^{1-\alpha}}ds$, $t\geq0$.

Let $t=0$ in (\ref{int_y_0}). Then, $0=y(0)=-\int_{0}^{+\infty}(Cy)(s)ds$. This means that we can recast the integral expression from (\ref{l1_issue_0}) as
\begin{eqnarray*}
y(t)=-C(t)\left(1-\int_{t}^{+\infty}y(s)ds\right)-\int_{t}^{+\infty}(Cy)(s)ds,\quad t\geq0,
\end{eqnarray*}
which is exactly (\ref{int_y_0}).

The proof is complete. $\square$

To give some insight to the (still unsettled) issue of {\lq\lq}$x^{\prime}\in L^{1}${\rq\rq}, notice that the condition $y(0)=0$ from Proposition \ref{prop1} reads as \begin{eqnarray*}
\int_{0}^{+\infty}x^{\prime}(s)\int_{0}^{s}\frac{a(\tau)}{(s-\tau)^{1-\alpha}}d\tau ds=0,
\end{eqnarray*}
which is really difficult to handle. A further intricacy is provided by the fact that, given $a\in C([0,+\infty),\mathbb{R})$, {\it the quantity $F(t)=\int_{0}^{t}\frac{\vert a(s)\vert}{(t-s)^{1-\alpha}}ds$, $t\geq0$, does not belong to $L^{1}([0,+\infty),\mathbb{R})$\/}. This follows from
\begin{eqnarray*}
\int_{T}^{t}F(2s)ds&\geq&\int_{T}^{t}\int_{\frac{T}{2}}^{2s}\frac{\vert a(\tau)\vert}{(2s-\tau)^{1-\alpha}}d\tau ds\geq\int_{T}^{t}\frac{ds}{\left(2s-\frac{T}{2}\right)^{1-\alpha}}\cdot\int_{\frac{T}{2}}^{2T}\vert a(\tau)\vert d\tau\\
&\rightarrow&+\infty\quad\mbox{when }t\rightarrow+\infty,
\end{eqnarray*}
where $T>0$ is chosen large enough for $a$ to be non-trivial in $\left[\frac{T}{2},2T\right]$.

\section{The case of ${}_{0}^{\>3}{\cal O}_{t}^{1+\alpha}$}
Introduce the relations
\begin{eqnarray}
y(t)=tx^{\prime}(t)-x(t),\quad x(t)=bt-t\int_{t}^{+\infty}\frac{y(\tau)}{\tau^{2}}d\tau,\quad t>0,\label{relint0}
\end{eqnarray}
with $b\neq0$ and $y\in{\cal RL}^{\alpha}((0,+\infty),\mathbb{R})$, see \cite{bma3}.

As before,
\begin{eqnarray*}
\frac{1}{\Gamma(1-\alpha)}\int_{0}^{t}\frac{y(s)}{(t-s)^{\alpha}}ds=x_{1}+\int_{t}^{+\infty}(ax)(s)ds,\quad t>0,
\end{eqnarray*}
where $x_{1}=\lim\limits_{t\rightarrow+\infty}\frac{1}{\Gamma(1-\alpha)}\int_{0}^{t}\frac{y(s)}{(t-s)^{\alpha}}ds\in\mathbb{R}$, and
\begin{eqnarray*}
&&\int_{0}^{t}y(s)ds\\
&&=\frac{x_{1}t^{\alpha}}{\Gamma(1+\alpha)}+\frac{1}{\Gamma(\alpha)}\int_{0}^{t}\frac{1}{(t-s)^{1-\alpha}}\int_{s}^{+\infty}(ax)(\tau)d\tau ds\\
&&=\frac{x_{1}t^{\alpha}}{\Gamma(1+\alpha)}+\frac{1}{\Gamma(\alpha)}\int_{0}^{t}\frac{1}{(t-s)^{1-\alpha}}\left(\int_{0}^{+\infty}-\int_{0}^{s}\right)(ax)(\tau)d\tau ds\\
&&=\frac{t^{\alpha}}{\Gamma(1+\alpha)}\left[x_{1}+\int_{0}^{+\infty}(ax)(\tau)d\tau\right]-\frac{1}{\Gamma(\alpha)}\int_{0}^{t}\int_{0}^{s}\frac{(ax)(u)}{(s-u)^{1-\alpha}}duds,
\end{eqnarray*}
see \cite[p. 32, eq. (2.13)]{SamkoKilbas}.

By differentiation, we get
\begin{eqnarray*}
y(t)=\frac{t^{\alpha-1}}{\Gamma(\alpha)}\left[x_{1}+\int_{0}^{+\infty}(ax)(\tau)d\tau\right]-\frac{1}{\Gamma(\alpha)}\int_{0}^{t}\frac{(ax)(s)}{(t-s)^{1-\alpha}}ds,
\end{eqnarray*}
where $t>0$.

Taking $a=-\frac{x_1}{(2-\alpha)\Gamma(\alpha)}$ and recalling (\ref{relint0}), our integral operator reads as
\begin{eqnarray*}
&&{\cal T}(y)(t)\\
&&=t^{\alpha-1}\left[a+\frac{b}{\Gamma(\alpha)}\int_{0}^{+\infty}sa(s)ds\right]-\frac{b}{\Gamma(\alpha)}\int_{0}^{t}\frac{sa(s)}{(t-s)^{1-\alpha}}ds\\
&&-\frac{t^{\alpha-1}}{\Gamma(\alpha)}\int_{0}^{+\infty}\tau a(\tau)\int_{\tau}^{+\infty}\frac{y(u)}{u^2}dud\tau\\
&&+\frac{1}{\Gamma(\alpha)}\int_{0}^{t}\frac{\tau a(\tau)}{(t-\tau)^{1-\alpha}}\int_{\tau}^{+\infty}\frac{y(u)}{u^2}dud\tau,\quad t>0.
\end{eqnarray*}

\begin{theorem}Assume that $\int_{0}^{+\infty}t\vert a(t)\vert dt+\sup\limits_{t>0}t^{1-\alpha}\int_{0}^{t}\frac{s\vert a(s)\vert}{(t-s)^{1-\alpha}}ds<+\infty$ and
\begin{eqnarray*}
\frac{1}{\Gamma(\alpha)}\left(\int_{0}^{+\infty}\frac{\vert a(s)\vert}{s^{1-\alpha}}ds+\chi\right)=k_{3}<1,
\end{eqnarray*}
where $\chi=\sup\limits_{t>0}t^{1-\alpha}\int_{0}^{t}\frac{\vert a(s)\vert}{(t-s)^{1-\alpha}s^{1-\alpha}}ds$. Then the FDE (\ref{gen_fde0}) for $i=3$ has a solution $x\in C^{1}((0,+\infty),\mathbb{R})$ with the asymptotic formula
\begin{eqnarray}
x(t)=[a+O(1)]t^{\alpha-1}+bt=bt+O(t^{\alpha-1})\quad\mbox{when }t\rightarrow+\infty.\label{as_form_x}
\end{eqnarray}
\end{theorem}

{\bf Proof.} Let us start by giving a simple example of $\chi$. If the functional coefficient $a\in (C\cap L^{1})([0,+\infty),\mathbb{R})$ verifies the restriction
\begin{eqnarray*}
\vert a(t)\vert\leq\frac{A}{t^{\alpha}},\quad t>0,
\end{eqnarray*}
then
\begin{eqnarray*}
&&t^{1-\alpha}\int_{0}^{2t}\frac{\vert a(s)\vert}{(2t-s)^{1-\alpha}s^{1-\alpha}}ds=t^{1-\alpha}\left(\int_{0}^{t}+\int_{t}^{2t}\right)\frac{\vert a(s)\vert}{(2t-s)^{1-\alpha}s^{1-\alpha}}ds\\
&&\leq t^{1-\alpha}\int_{0}^{t}\frac{\vert a(s)\vert}{t^{1-\alpha}s^{1-\alpha}}ds+t^{1-\alpha}\int_{t}^{2t}\frac{A}{(2t-s)^{1-\alpha}s}ds\\
&&\leq\left(\int_{0}^{1}\frac{\vert a(s)\vert}{s^{1-\alpha}}ds+\int_{1}^{1+t}\frac{\vert a(s)\vert}{s^{1-\alpha}}ds\right)+A\int_{\frac{1}{2}}^{1}\frac{dv}{(1-v)^{1-\alpha}v}\\
&&\leq\left(\int_{0}^{1}\frac{ds}{s^{1-\alpha}}\cdot\Vert a\Vert_{L^{\infty}(0,1)}+\int_{1}^{+\infty}\vert a(s)\vert ds\right)+2A\int_{\frac{1}{2}}^{1}\frac{dv}{(1-v)^{1-\alpha}}\\
&&\leq\frac{1}{\alpha}\Vert a\Vert_{L^{\infty}(0,1)}+\Vert a\Vert_{L^{1}(1,+\infty)}+A\frac{2^{1-\alpha}}{\alpha}<+\infty,\quad t>0.
\end{eqnarray*}

Notice also that $\int_{0}^{t}\frac{\vert a(s)\vert}{(t-s)^{1-\alpha}}ds\leq t^{1-\alpha}\int_{0}^{t}\frac{\vert a(s)\vert}{(t-s)^{1-\alpha}s^{1-\alpha}}ds\leq\chi$ and
\begin{eqnarray*}
t^{1-\alpha}\int_{0}^{t}\frac{s\vert a(s)\vert}{(t-s)^{1-\alpha}}ds=t^{1-\alpha}\int_{0}^{t}\frac{s^{2-\alpha}\vert a(s)\vert}{(t-s)^{1-\alpha}s^{1-\alpha}}ds,\quad t>0,
\end{eqnarray*}
which leads to the {\lq\lq}$\chi${\rq\rq} of the mapping $t\mapsto t^{2-\alpha}a(t)$ in $[0,+\infty)$.

Introduce now the set $Z$ of all the functions $y\in C((0,+\infty),\mathbb{R})$ such that $\sup\limits_{t>0}t^{1-\alpha}\vert y(t)\vert<+\infty$ and the metric
\begin{eqnarray*}
d(y_{1},y_{2})=\sup\limits_{t>0}t^{1-\alpha}\vert y_{1}(t)-y_{2}(t)\vert,\quad y_{1},y_{2}\in Z.
\end{eqnarray*}
Observe also that
\begin{eqnarray}
\sup\limits_{t>0}t^{2-\alpha}\int_{t}^{+\infty}\frac{\vert y_{1}(u)-y_{2}(u)\vert}{u^{2}}du&\leq&\frac{1}{2-\alpha}\cdot\sup\limits_{t>0}t^{1-\alpha}\vert y_{1}(t)-y_{2}(t)\vert\nonumber\\
&\leq&d(y_{1},y_{2}).\label{tehn0}
\end{eqnarray}

The metric space ${\cal P}=(Z,d)$ is complete. Given $y\in {\cal P}$, we have the estimates
\begin{eqnarray*}
&&t^{1-\alpha}\vert{\cal T}(y)(t)\vert\\
&&\leq\vert a\vert+\frac{\vert b\vert}{\Gamma(\alpha)}\int_{0}^{+\infty}s\vert a(s)\vert ds+\frac{\vert b\vert}{\Gamma(\alpha)}\cdot\sup\limits_{t>0}t^{1-\alpha}\int_{0}^{t}\frac{s\vert a(s)\vert}{(t-s)^{1-\alpha}}ds\\
&&+\frac{1}{\Gamma(\alpha)}\int_{0}^{+\infty}\frac{\vert a(s)\vert}{s^{1-\alpha}}ds\cdot\sup\limits_{s>0}s^{2-\alpha}\int_{s}^{+\infty}\frac{\vert y(u)\vert}{u^2}du\\
&&+\frac{1}{\Gamma(\alpha)}\cdot\sup\limits_{t>0}t^{1-\alpha}\int_{0}^{t}\frac{\vert a(\tau)\vert}{(t-\tau)^{1-\alpha}\tau^{1-\alpha}}d\tau\cdot\sup\limits_{\tau>0}\tau^{2-\alpha}\int_{\tau}^{+\infty}\frac{\vert y(u)\vert}{u^2}du,\quad t>0,
\end{eqnarray*}
which imply that ${\cal T}({\cal P})\subseteq{\cal P}$.

Further, we have
\begin{eqnarray*}
&&t^{1-\alpha}\vert{\cal T}(y_{1})(t)-{\cal T}(y_{2})(t)\vert\\
&&\leq\left[\frac{1}{\Gamma(\alpha)}\int_{0}^{+\infty}\frac{\vert a(s)\vert}{s^{1-\alpha}}ds+\frac{1}{\Gamma(\alpha)}\sup\limits_{t>0}t^{1-\alpha}\int_{0}^{t}\frac{\vert a(\tau)\vert}{(t-\tau)^{1-\alpha}\tau^{1-\alpha}}d\tau\right]\\
&&\times d(y_{1},y_{2})\\
&&=\frac{1}{\Gamma(\alpha)}\left(\int_{0}^{+\infty}\frac{\vert a(s)\vert}{s^{1-\alpha}}ds+\chi\right)d(y_{1},y_{2}),\quad t>0,
\end{eqnarray*}
by means of (\ref{tehn0}), where $y_{1}$, $y_{2}\in{\cal P}$.

The operator ${\cal T}:{\cal P}\rightarrow{\cal P}$ being a contraction of coefficient $k_{3}$, it has a fixed point $y_{0}$. Thus, since $y_{0}(t)=O(t^{\alpha-1})$ for large values of $t$, we conclude the validity of the asymptotic expansion (\ref{as_form_x}) for the solution $x$ given by (\ref{relint0}). Notice also that
\begin{eqnarray*}
\lim\limits_{t\searrow0}t^{1-\alpha}y_{0}(t)=\lim\limits_{t\searrow0}t^{1-\alpha}{\cal T}(y_{0})(t)=a+\frac{b}{\Gamma(\alpha)}\int_{0}^{+\infty}sa(s)ds.
\end{eqnarray*}

The proof is complete. $\square$

\section{The case of ${}_{0}^{\>2}{\cal O}_{t}^{1+\alpha}$}
The asymptotic formula (\ref{as_form_gen}) has been already discussed in \cite{bma2}, however, it is worthy to be recalled for reasons of completeness.

\begin{theorem}
(\cite[Theorem 1]{bma2}) Assume that there exists $T>0$ such that
\begin{eqnarray*}
\frac{\max\{1,T\}}{\Gamma(1+\alpha)}\left(\int_{0}^{T}\frac{\vert a(s)\vert}{s^{1+\alpha}}ds+\int_{T}^{+\infty}s^{\alpha}\vert a(s)\vert ds\right)=k_{4}<1
\end{eqnarray*}
and $\int_{T}^{+\infty}s^{1+\alpha}\vert a(s)\vert ds<+\infty$. Then, given $a$, $b\in\mathbb{R}$, with $a^{2}+b^{2}>0$, the FDE (\ref{gen_fde0}) for $i=2$ has a solution $x\in C((0,+\infty),\mathbb{R})$ with the asymptotic formula
\begin{eqnarray*}
x(t)=[a+O(1)]t^{\alpha-1}+bt^{\alpha}=bt^{\alpha}+O(t^{\alpha-1})\quad\mbox{when }t\rightarrow+\infty.
\end{eqnarray*}
\end{theorem}

The formula of the integral operator reads in this case as
\begin{eqnarray*}
&&{\cal T}(x)(t)=at^{\alpha-1}+bt^{\alpha}+\frac{1}{\Gamma(\alpha)}\int_{0}^{t}\frac{1}{(t-s)^{1-\alpha}}\int_{s}^{+\infty}(ax)(\tau)d\tau ds,\quad t>0,
\end{eqnarray*}
and its fixed point $x_{0}$ satisfies also the conditions
\begin{eqnarray*}
\lim\limits_{t\searrow0}t^{1-\alpha}x_{0}(t)=a,\quad\lim\limits_{t\rightarrow+\infty}({}_{0}D_{t}^{\alpha}x_{0})(t)=\Gamma(1+\alpha)b.
\end{eqnarray*}

\end{document}